\documentclass[
 reprint,
 superscriptaddress,
 amsmath,amssymb,
 aps,
 prl,
]{revtex4-1}

\usepackage{graphicx}
\usepackage{dcolumn}
\usepackage{bm}
\usepackage{braket}
\usepackage{hyperref}
\usepackage{natbib}
\usepackage{upgreek}
\hypersetup{colorlinks=true,linkcolor=blue,citecolor=blue,urlcolor=blue}

\begin{document}

\preprint{Preprint ID}

\title{Coherent optical control of a quantum-dot spin-qubit in a waveguide-based spin-photon interface}

\author{Dapeng Ding}
\author{Martin Hayhurst Appel}
\author{Alisa Javadi}
\author{Xiaoyan Zhou}
\affiliation{Center for Hybrid Quantum Networks (Hy-Q), Niels Bohr Institute, University of Copenhagen, Blegdamsvej 17, DK-2100 Copenhagen, Denmark}
\author{Matthias Christian L\"{o}bl}
\author{Immo S\"{o}llner}
\affiliation{Department of Physics, University of Basel, Basel, Switzerland}
\author{R\"{u}diger Schott}
\affiliation{Lehrstuhl f\"{u}r Angewandte Festk\"{o}rperphysik, Ruhr-Universit\"{a}t Bochum, Bochum, Germany}
\author{Camille Papon}
\author{Tommaso Pregnolato}
\author{Leonardo Midolo}
\affiliation{Center for Hybrid Quantum Networks (Hy-Q), Niels Bohr Institute, University of Copenhagen, Blegdamsvej 17, DK-2100 Copenhagen, Denmark}
\author{Andreas Dirk Wieck}
\affiliation{Lehrstuhl f\"{u}r Angewandte Festk\"{o}rperphysik, Ruhr-Universit\"{a}t Bochum, Bochum, Germany}
\author{Arne Ludwig}
\affiliation{Lehrstuhl f\"{u}r Angewandte Festk\"{o}rperphysik, Ruhr-Universit\"{a}t Bochum, Bochum, Germany}
\author{Richard John Warburton}
\affiliation{Department of Physics, University of Basel, Basel, Switzerland}
\author{Tim Schr\"{o}der}
\author{Peter Lodahl}
\email{lodahl@nbi.ku.dk}
\affiliation{Center for Hybrid Quantum Networks (Hy-Q), Niels Bohr Institute, University of Copenhagen, Blegdamsvej 17, DK-2100 Copenhagen, Denmark}

\date{\today}

\begin{abstract}
Waveguide-based spin-photon interfaces on the GaAs platform have emerged as a promising system for a variety of quantum information applications directly integrated into planar photonic circuits. The coherent control of spin states in a quantum dot can be achieved by applying circularly polarized laser pulses that may be coupled into the planar waveguide vertically through radiation modes. However, proper control of the laser polarization is challenging since the polarization is modified through the transformation from the far field to the exact position of the quantum dot in the nanostructure. Here we demonstrate polarization-controlled excitation of a quantum-dot electron spin and use that to perform coherent control in a Ramsey interferometry experiment. The Ramsey interference reveals a pure dephasing time of $ 2.2\pm0.1 $~ns, which is comparable to the values so far only obtained in bulk media. We analyze the experimental limitations in spin initialization fidelity and Ramsey contrast and identify the underlying mechanisms.

\begin{description}
\item[DOI]
\end{description}

\end{abstract}


\maketitle

Stationary spin qubits coupled to coherent photons are the basis for a variety of quantum information applications including the implementation of quantum gates \cite{duan_scalable_2004,hacker_photonphoton_2016,welte_photon-mediated_2018,li_quantum_2018}, the generation of photonic cluster states \cite{lindner_proposal_2009,schwartz_deterministic_2016}, and the construction of quantum networks \cite{nemoto_photonic_2016,mahmoodian_quantum_2016,lodahl_quantum-dot_2018}. These applications require efficient spin-photon interfaces \cite{yao_theory_2005,yilmaz_quantum-dot-spin_2010} and their integration into photonic systems that allow for the coherent control of individual qubits \cite{press_complete_2008} as well as the generation of spin-photon \cite{wilk_single-atom_2007,togan_quantum_2010,mi_coherent_2018} and spin-spin entanglement \cite{ritter_elementary_2012,de_greve_quantum-dot_2012,kalb_entanglement_2017}. Spin-photon interfaces can be realized by coupling a spin to an optical cavity \cite{sun_single-photon_2018}, or alternatively, to an optical waveguide \cite{luxmoore_interfacing_2013,sollner_deterministic_2015}. Waveguide systems are advantageous in terms of reduced fabrication complexity, broadband operation, chiral mode coupling \cite{sollner_deterministic_2015,price_non-reciprocal_2018}, and direct integration into complex photonic circuits \cite{mouradian_scalable_2015,davanco_heterogeneous_2017}. Waveguide-based photonic circuits offer functionalities ranging from beam splitters \cite{prtljaga_monolithic_2014}, over fast switching \cite{midolo_electro-optic_2017}, to single-photon detectors \cite{sahin_waveguide_2015,vetter_cavity-enhanced_2016}, and enable, among others, quantum nonlinear optics \cite{bhaskar_quantum_2017,javadi_single-photon_2015} and multi-qubit entanglement generation \cite{sipahigil_integrated_2016}.

One of the most promising material platforms for combining stationary qubits with classical control functionalities in photonic integrated circuits is gallium arsenide (GaAs). Waveguide-integrated indium gallium arsenide (InGaAs) quantum dots constitute excellent spin-photon interfaces with photon coupling efficiencies ($\beta$-factor) of $ > $98\% \cite{arcari_near-unity_2014}, near-lifetime-limited single-photon emission \cite{thyrrestrup_quantum_2018,liu_high_2018}, multi-photon probability as low as $10^{-4}$ \cite{hanschke_quantum_2018}, and access to quasi-permanent spin qubits with near-unity state preparation fidelities \cite{javadi_spinphoton_2018}. Furthermore, the generation of spin-polarized excitons \cite{luxmoore_optical_2013} and spin-state-controlled photon switching \cite{javadi_spinphoton_2018} have been demonstrated.

A key functionality for many quantum applications is the ability to prepare a coherent superposition of the two spin eigenstates $\ket{\uparrow}$ and $\ket{\downarrow}$. Such a state may be prepared using circularly polarized laser pulses \cite{press_ultrafast_2010}. In nanophotonic devices this is an experimental challenge since the polarization of the laser pulses changes when it is coupled from the far field into the photonic nanostructure. The proper control of the laser polarization is therefore essential. In addition, nanostructures may deteriorate coherence properties of the quantum dot through surface defect states or modified phonon modes \cite{tighineanu_phonon_2018}.

In this article, we report on the coherent optical control of a quantum-dot electron spin embedded in a nanobeam waveguide. We show that by sensitively controlling the laser polarization it is possible to drive a circularly polarized transition of the quantum dot. This polarization setting is used to demonstrate Ramsey interference with an extracted pure dephasing time $ T_2^* $ of $ 2.2\pm0.1 $~ns, which is limited by the coupling to the fluctuating nuclear spin bath and could potentially be extended by narrowing its distribution \cite{ethier-majcher_improving_2017}. The observed coherence time is comparable to the value reported for quantum dots in bulk media \cite{bechtold_three-stage_2015,stockill_quantum_2016,ethier-majcher_improving_2017} and extends previous work on quantum-dot electron spins in photonic-crystal cavities  \cite{carter_quantum_2013,sun_quantum_2016}.

\begin{figure}[htb]
\centering
\includegraphics[width=86mm]{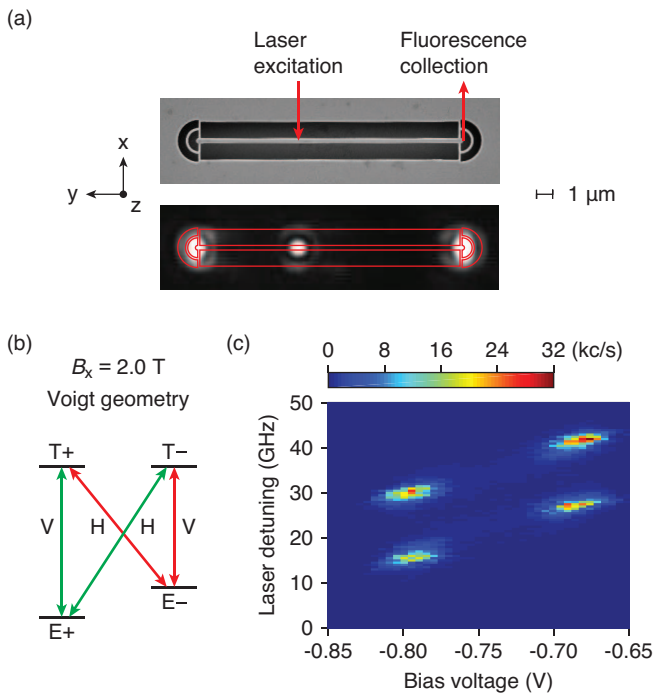}
\caption{Microscope images, energy level diagram, and plateau map in the Voigt geometry. (a) Scanning electron microscope image (top) and fluorescence microscope image overlaid with the design layout of the device (bottom). An InGaAs quantum dot is embedded in a GaAs nanobeam waveguide. The waveguide is coupled to a circular grating coupler at each end. The excitation lasers are focused on the quantum dot and the fluorescence of the quantum dot is collected at one of the grating couplers. (b) Energy level diagram of a quantum dot under a magnetic field in the Voigt geometry ($ B_x=2.0 $~T). The ground states $ \ket{E \pm} $ are electron-spin states. The excited states $ \ket{T \pm} $ are trion states consisting of two electrons and a heavy hole. Four optical transitions are coupled to linearly polarized light with vertical (V) and horizontal (H) polarizations. (c) Resonance fluorescence intensity as a function of laser detuning and bias voltage in the Voigt geometry. The central dark region occurs due to optical pumping of the spin. The Zeeman splitting of the excited state is close to the linewidth of the trion transitions and thus only two optical transitions are resolved.}
\label{fig:device}
\end{figure}

We explore the spin states of a single electron in an InGaAs quantum dot embedded in a nanobeam waveguide. This device has been studied in Ref.~\cite{javadi_spinphoton_2018} where the detailed description and characterization can be found. All experiments in the present work were conducted on the same quantum dot. A scanning electron microscope image of the device is shown in Fig.~\hyperref[fig:device]{\ref{fig:device}(a)}. It is fabricated from a 175 nm thick membrane suspended above a GaAs substrate. The membrane is grown by molecular beam epitaxy with multiple intrinsic (I), $ p $-doped (P), and $ n $-doped (N) GaAs layers forming a PININ diode structure. A quantum-dot layer is grown in the middle of the membrane along the growth direction ($ z $ axis). This structure ensures a modest electric field variation across the quantum dots \cite{lobl_narrow_2017}. The waveguide is 300 nm wide and 16 $ \mu $m long. It is coupled to a circular grating coupler at each end. In the experiment, the device is mounted in a closed-cycle cryostat at 4 K with optical access along the $ z $ axis. The excitation laser light is focused directly on the quantum dot from the top of the sample by coupling through the radiation modes of the waveguide. Subsequently, the fluorescence from the quantum dot is coupled to the waveguide and collected by one of the grating couplers. Figure~\hyperref[fig:device]{\ref{fig:device}(a)} shows a fluorescence microscope image of the device subject to strong excitation at 830~nm wavelength. Fluorescence from the quantum dot directly and diffracted by the two grating couplers is clearly visible.

Coherent control (rotation) of quantum-dot spin states in a bulk medium has been demonstrated using ultrafast laser pulses in the Voigt geometry where a magnetic field is oriented in the sample plane (the $xy$ plane here) \cite{press_complete_2008}. An energy level diagram of a quantum dot under a magnetic field in the Voigt geometry ($ B_x=2.0 $ T) is shown in Fig.~\hyperref[fig:device]{\ref{fig:device}(b)}. Four optical transitions are coupled to linearly polarized light. The ground states are the superposition of the eigenstates in the original basis (the $ z $ axis) for the quantum dot at zero magnetic field such that $ \ket{E \pm}=(\ket{\uparrow}\pm \ket{\downarrow})/\sqrt{2} $. Similarly, the excited states (trion) are given by $ \ket{T \pm}=(\ket{\uparrow \downarrow \Uparrow}\pm \ket{\downarrow \uparrow \Downarrow})/\sqrt{2} $, where $ \ket{\Uparrow} $ and $ \ket{\Downarrow} $ are the eigenstates of a heavy hole.

To determine the level structure, we measure the fluorescence intensity of the quantum dot as a function of laser detuning and bias voltage. A narrow-linewidth continuous-wave laser is used to excite the quantum dot with a linear polarization along the waveguide (the $ y $ axis) for the best extinction of the laser background. The result, which is typically referred to as a plateau map, is shown in Fig.~\hyperref[fig:device]{\ref{fig:device}(c)}. The voltage range approximately from $-0.79$ V to $-0.68$ V corresponds to the single-electron-charged states (plateaus). The two plateaus are separated by $ \sim $14.5~GHz primarily due to the Zeeman splitting of the ground states with an electron $ g $-factor of $ -0.5 $. The Zeeman splitting of the excited states is close to the linewidth of the trion transitions and thus cannot be resolved in the plateau map. Between the plateau edges, optical pumping occurs resulting in weak fluorescence as the electron is prepared in the state that is not resonant with the laser. The optical pumping fidelity is 90\% in the limit of strong excitation. At the edges of the plateaus, strong cotunneling with the back contact of the diode prevents optical pumping resulting in strong fluorescence seen as four bright regions in Fig.~\hyperref[fig:device]{\ref{fig:device}(c)} \cite{smith_voltage_2005}.

\begin{figure}[htb]
\centering
\includegraphics[width=86mm]{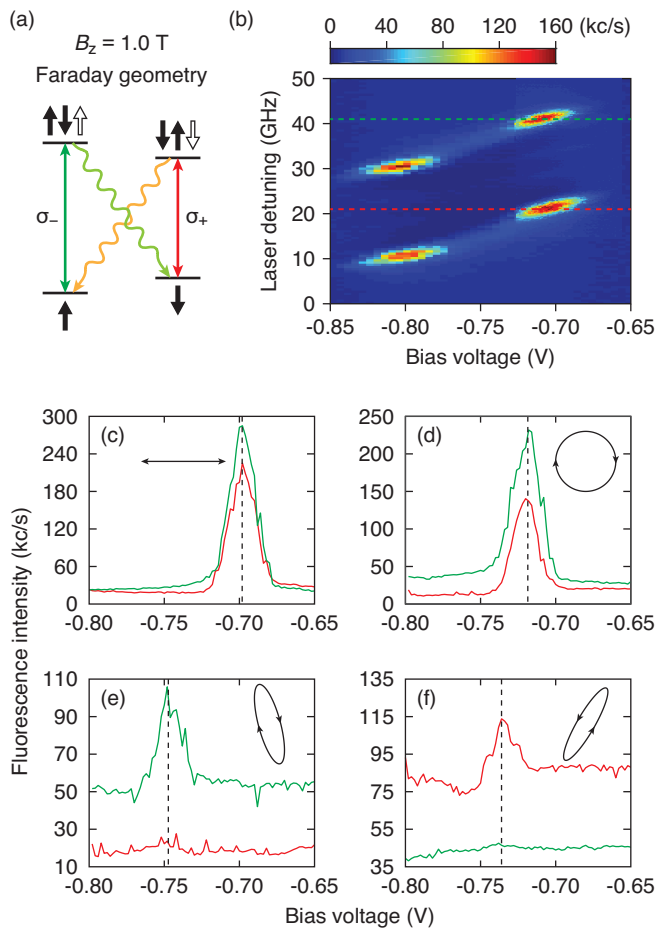}
\caption{Energy level diagram, plateau map, and resonance fluorescence in the Faraday geometry. (a) Energy level diagram of a  quantum dot under a magnetic field in the Faraday geometry ($ B_z=1.0 $~T). The ground states are coupled to the corresponding excited states via circularly polarized transitions with opposite helicity (vertical lines). The two diagonal transitions (wavy lines) are weakly allowed. (b) Resonance fluorescence intensity as a function of laser detuning and bias voltage in the Faraday geometry. The resonance frequencies at the edges of the plateaus are determined, as indicated by red and green dashed lines corresponding to the two circular dipoles, as shown in (a) with the same color notations. (c)--(f) Resonance fluorescence intensity as a function of bias voltage for the two circular transitions with (c) linear, (d) circular, and (e) and (f) elliptic laser polarizations. The laser frequency of the green (red) curve corresponds to the green (red) dashed line in (b). The resonance voltages are indicated by dashed lines. The polarization in (f) is used for Ramsey interference measurements.}
\label{fig:dipole}
\end{figure}

The rotation of spin states of a quantum dot using ultrafast laser pulses can be described in terms of an AC Stark shift \cite{berezovsky_picosecond_2008}. In this picture, a circularly polarized rotation laser pulse drives only one of the transitions in the original basis resulting in an energy shift between the two ground states. This energy shift is equivalent to a rotation in the hybridized basis defined in the Voigt geometry. Therefore circularly polarized laser pulses can effectively rotate the spin states of a quantum dot. The situation is complicated when the quantum dot is embedded in a waveguide. In general, a transition dipole, which is well coupled to the waveguide mode, is weakly coupled to free-space modes and the polarization transformation when coupling through radiation modes depends on the spatial position of the quantum dot. The Supplemental Material presents numerical simulations of the polarization transformation. At some particular positions, pure chiral coupling to the waveguide is possible \cite{coles_chirality_2016}. Consequently, selecting the polarization in the far field in order to precisely excite a certain quantum dot transition is a non-trivial task.

In order to determine the far-field polarization that is required to excite the circular dipoles in the waveguide, it is convenient to operate first in the Faraday geometry where a magnetic field is applied along the growth direction. In the Faraday geometry, the same selection rules apply as in the original basis while the degeneracy of the two circular dipoles is lifted by the magnetic field. An energy level diagram of a quantum dot under a magnetic field in the Faraday geometry ($ B_z=1.0 $~T) is shown in Fig.~\hyperref[fig:dipole]{\ref{fig:dipole}(a)}. The ground states are coupled to the corresponding excited states via circularly polarized transitions with opposite helicity (vertical lines). The two diagonal transitions (wavy lines) are weakly allowed due to light-heavy hole mixing and hyperfine interactions \cite{dreiser_optical_2008}. The measured plateau map in the Faraday geometry with $ B_z=1.0 $~T is shown in Fig.~\hyperref[fig:dipole]{\ref{fig:dipole}(b)}. We determine the resonance frequencies at the edges of the plateaus (dashed lines), corresponding to the two circular dipoles, as shown in Fig.~\hyperref[fig:dipole]{\ref{fig:dipole}(a)}. These two frequencies will be used to study the far-field polarization required to excite the two circular dipoles.

In the experiment, a narrow-linewidth laser is tuned to the two frequencies, respectively, and resonance fluorescence intensity as a function of bias voltage is recorded. The ellipticity and orientation of the polarization of the laser are scanned by a combination of a half-wave plate and a quarter-wave plate on motorized rotation stages. The measured fluorescence intensities as a function of bias voltage for four different polarizations are shown in Figs.~\hyperref[fig:dipole]{\ref{fig:dipole}(c)}--\hyperref[fig:dipole]{\ref{fig:dipole}(f)} (see Supplemental Material for complete polarization space). In Figs.~\hyperref[fig:dipole]{\ref{fig:dipole}(c)} and \hyperref[fig:dipole]{\ref{fig:dipole}(d)}, linear and circular polarizations are used, respectively, resulting in excitation of both of the two circular dipoles. We find that at two particular elliptic polarizations the contrast between the two dipoles is maximized as shown in Figs.~\hyperref[fig:dipole]{\ref{fig:dipole}(e)} and \hyperref[fig:dipole]{\ref{fig:dipole}(f)} with a ratio of about ten. For these two polarizations the excitation of one of the circular dipoles is suppressed indicating that the laser polarization is orthogonal to this dipole. The polarization in Fig.~\hyperref[fig:dipole]{\ref{fig:dipole}(f)} will be used for coherent control of the spin states.

\begin{figure}[htb]
\centering
\includegraphics[width=86mm]{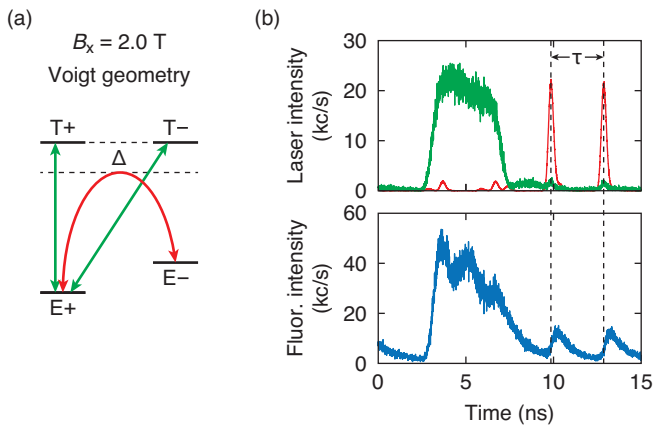}
\caption{Ramsey interference experiment of an electron spin in a quantum dot embedded in a nanobeam waveguide. (a) Energy level diagram of a quantum dot under a magnetic field in the Voigt geometry. The in-plane $ g $-factor of the hole of this particular quantum dot is nearly zero and thus the excited state is almost degenerate. A narrow-linewidth laser drives two optical transitions (green arrowed lines) and performs optical pumping for state initialization and readout. Red-detuned ($ \Delta=-0.8 $ THz) laser pulses are used to rotate the ground states (red arrowed curve). (b) Time traces of laser pulses (top) and fluorescence intensity (bottom) in the measurement of the Ramsey interference. The resonant laser pulse (green) is $ \sim $5 ns long. A pair of rotation laser pulses of $ \sim $6 ps in width is separated by a variable delay time $ \tau $. The polarization of the rotation pulses is the same as in Fig.~\hyperref[fig:dipole]{\ref{fig:dipole}(d)}. The fluorescence intensity during the resonant laser pulses decreases due to optical pumping.}
\label{fig:ramsey_scheme}
\end{figure}

\begin{figure*}[htb]
\centering
\includegraphics[width=179mm]{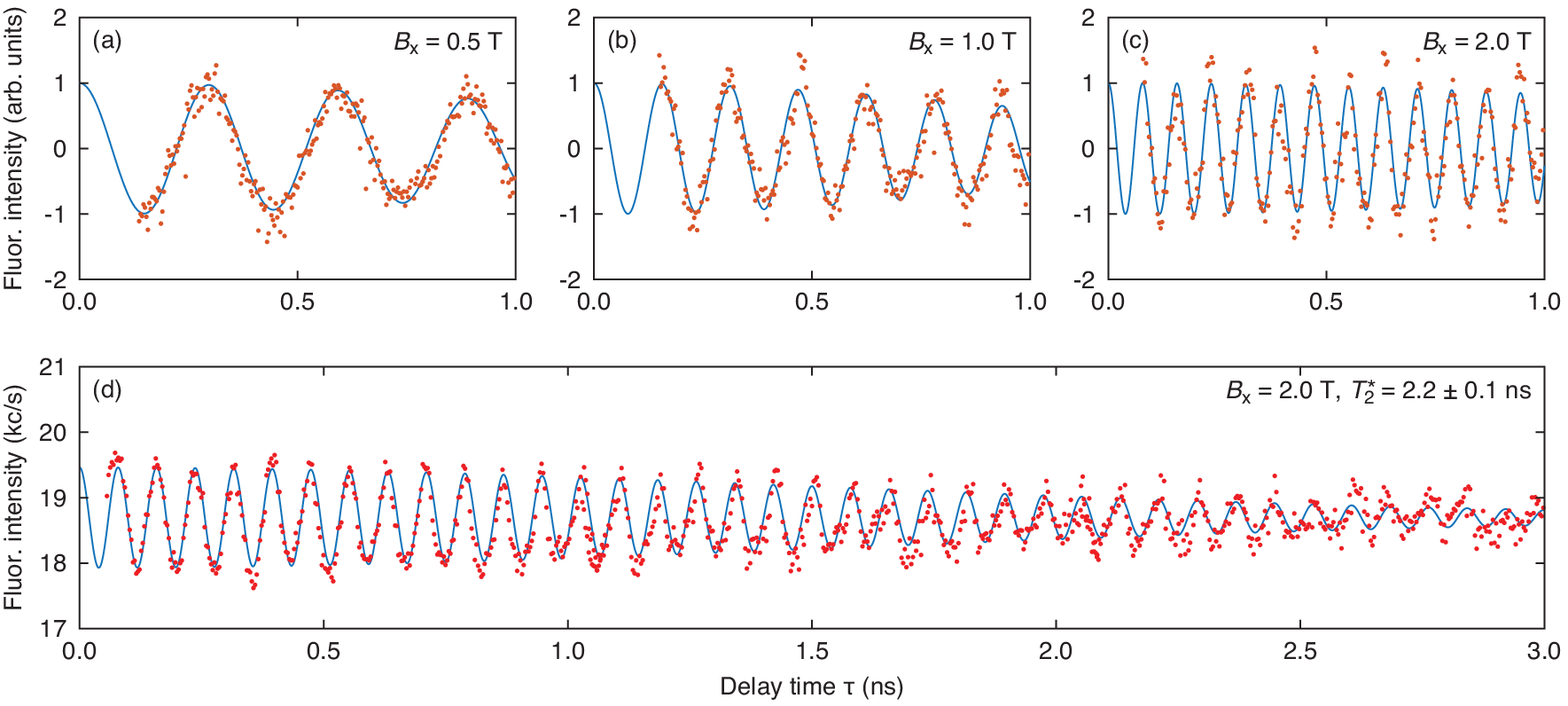}
\caption{Ramsey interference of an electron spin in a quantum dot embedded in a nanobeam waveguide. Fluorescence intensities as a function of delay time $ \tau $ between the two rotation pulses at (a) 0.5, (b) 1.0 T, and (c) and (d) 2.0 T magnetic fields. In (a)--(c), the background has been subtracted and the intensity is centered around zero, while in (d) the raw data exhibit a contrast of 0.04 for the first oscillation period. In (d) three sets of data are separately measured and combined due to limited length of the optical delay line. The fit model is a cosine function with a Gaussian envelope. The time constant of the Gaussian function is a measure of the pure dephasing time $ T_2^* $ of the spin states of the electron. The extracted value of $ T_2^* $ is $ 2.2\pm0.1 $~ns.}
\label{fig:ramsey}
\end{figure*}

For the demonstration of coherent control of the spin states through the Ramsey interference, we switch to the Voigt geometry. Figure~\hyperref[fig:ramsey_scheme]{\ref{fig:ramsey_scheme}(a)} shows the energy level diagram and the laser schemes. A narrow-linewidth laser resonantly drives the two higher-energy transitions and performs optical pumping for state initialization and readout. A red-detuned ($ \Delta=-0.8 $ THz) laser is used to rotate the spin states with the polarization shown in the inset of Fig.~\hyperref[fig:dipole]{\ref{fig:dipole}(f)}. The laser pulse sequence for the Ramsey experiment is shown in Fig.~\hyperref[fig:ramsey_scheme]{\ref{fig:ramsey_scheme}(b)}. A pair of rotation laser pulses of $ \sim $6 ps in width is separated by a variable delay time $ \tau $. The laser power is calibrated for a rotation of $ \pi/2 $ in the Bloch sphere (see Supplemental Material for the calibration). The resonant laser pulse is $ \sim $5 ns long. The measured fluorescence intensity as a function of time is shown in Fig.~\hyperref[fig:ramsey_scheme]{\ref{fig:ramsey_scheme}(b)}. The fluorescence intensity decreases during the resonant laser pulse due to optical pumping, but it does not completely decay implying a low optical pumping fidelity. It exhibits Rabi oscillations, which indicates that the Rabi frequency is larger than the optical pumping rate. The optical pumping fidelity is limited by the ratio of the excited-state decay rate to the ground-state spin-flip rate and also by off-resonant repumping via the lower-energy transitions. Furthermore, the red-detuned laser pulses do not only rotate the ground states, but also excite the trion states, as visible in Fig.~\hyperref[fig:ramsey_scheme]{\ref{fig:ramsey_scheme}(b)} around 10 and 13~ns.

Figure~\ref{fig:ramsey} shows the result of the Ramsey experiment. In Figs.~\hyperref[fig:ramsey]{\ref{fig:ramsey}(a)}--\hyperref[fig:ramsey]{\ref{fig:ramsey}(c)}, resonance fluorescence intensity (red dots) is measured as a function of delay time $ \tau $ at $ B_x=0.5 $, 1.0, and 2.0 T, respectively. The background has been subtracted and the intensity is centered around zero. The fit model (blue curves) is a cosine function with a Gaussian envelope. The extracted Larmor frequencies ($ \omega_\text{L}/2\pi $) are $ 3.37\pm0.01 $, $ 6.41\pm0.01 $, and $ 12.70\pm0.02 $ GHz at 0.5, 1.0, and 2.0 T, corresponding to the $ g $-factor of $ -0.48 $, $ -0.46 $, and $ -0.45 $, respectively. The slight variation is probably due to the uncertainty of the magnetic field strength since the sample may not be exactly located at the position where the magnetic field is calibrated. The Ramsey interference trace at 2.0~T at long delay times is shown in Fig.~\hyperref[fig:ramsey]{\ref{fig:ramsey}(d)}. The plot is constructed from three separately measured data sets due to the limited length of the optical delay line. The fit yields a pure dephasing time $ T_2^\ast=2.2\pm0.1 $ ns for the quantum dot in a nanobeam waveguide, which is similar to the typical value found in bulk media \cite{bechtold_three-stage_2015,stockill_quantum_2016,ethier-majcher_improving_2017}.

The demonstration of coherent spin-state rotations and bulk-like pure dephasing time indicates that quantum-dot-nanobeam-waveguide systems are promising spin-photon interfaces. However, in the present experimental implementation a limited Ramsey contrast of $C \approx 0.04$ is observed, cf. Fig.~\hyperref[fig:ramsey]{\ref{fig:ramsey}(d)}, where $ C = (I_\text{max}-I_\text{min})/(I_\text{max}+I_\text{min}) $, with $ I_\text{max} $ and $ I_\text{min} $ being the maximum and minimum intensities in the first oscillation period, respectively. We attribute this limitation to the significantly higher laser power ($ \sim $25 $ \mu $W mean power on the quantum dot) required for a $ \pi/2 $ rotation compared to that for quantum dots in bulk media with a similar frequency detuning $ \Delta $. A high rotation laser power is required due to the high coupling efficiency of the quantum dot to the waveguide with a $ \beta $-factor of $ \sim $80\% and therefore low coupling efficiency to the laser in free space.

We identify several mechanisms, through which the rotation laser can lead to a reduced Ramsey contrast. During the coherent control sequence, the rotation laser adversely excites the trion population as seen in Fig.~\hyperref[fig:ramsey_scheme]{\ref{fig:ramsey_scheme}(b)}. Populating the excited states directly reduces the rotation fidelity. More seriously, the rotation laser also creates free charge carriers in the waveguide. In the experiment, the resonance voltage of the quantum dot transitions shifts from $ -0.74 $ V without the rotation laser (Fig.~\hyperref[fig:dipole]{\ref{fig:dipole}(d)}) to $ -1.41 $ V with the rotation laser, which indicates that a more negative bias voltage is needed to compensate the internal electric field built by the free charge carriers. The linewidth in bias voltage is also broadened from 0.02 V without the rotation laser (Fig.~\hyperref[fig:dipole]{\ref{fig:dipole}(d)}) to 0.1 V with the rotation laser (see Supplemental Material for details). The increased linewidth leads to an increased repumping via the lower-energy transitions. At the same time, the excess free charge carriers are likely to cause an increase in the spin-flip rate of the quantum dot. Both the line broadening and increased spin-flip rate reduce the optical pumping fidelity and are, together with the trion excitation by the rotation laser, responsible for the low contrast of the Ramsey interference.

We simulate the dynamics of the quantum dot using a four-level model and take these three effects into account. By matching the fluorescence time trace (Fig.~\hyperref[fig:ramsey_scheme]{\ref{fig:ramsey_scheme}(b)}) and the Ramsey contrast we obtain an initialization fidelity of 54\% and a spin-flip rate of 90 $ \mu $s$ ^{-1} $. In contrast, the fidelity is 90\% in the absence of the rotation laser and a spin-flip rate of 0.2 $ \mu $s$ ^{-1} $ has been observed on the same quantum dot  \cite{javadi_spinphoton_2018}. Despite a low initialization fidelity, we obtain a high rotation fidelity of 99\% from the model (see Supplemental Material for details).

We would like to point out possible methods to mitigate the experimental limitations induced by the rotation laser. In our device, the top $ p $-doped layer was over-etched during the fabrication resulting in a thin layer with a large resistance, which is less efficient for suppressing charge noise in the device. As a result, the linewidth of the trion transitions of 1.8 GHz is significantly larger than the lifetime-limited value of $ \sim $0.2 GHz. For the same reason the excess free charge carriers created by the rotation laser could not be efficiently removed. We note that in our next-generation devices with improved designs lifetime-limited linewidth was achieved \cite{thyrrestrup_quantum_2018}. We anticipate much less detrimental effects of the rotation laser on these devices. Finally, we propose to couple the rotation laser through the waveguide for a chirally-coupled quantum dot \cite{coles_chirality_2016}. In this way the laser field can efficiently interact with the circular dipoles of the quantum dot, reducing the required laser power.

In summary, we have investigated an electron spin in a quantum dot as a stationary qubit. The spin degrees of freedom are efficiently coupled to photons via a spin-photon interface in a nanobeam waveguide. The qubit state is controlled by laser pulses with a pre-determined free-space polarization required for optically accessing the quantum dot in the waveguide-modified dielectric environment. We determine the required polarization by mapping out the free-space to waveguide polarization transformation. This method can be directly applied to other photonic structures such as fibre tapers and photonic-crystal waveguides and cavities. We subsequently use this polarization to demonstrate Ramsey interference. The extracted pure dephasing time of $ 2.2\pm0.1 $~ns is similar to quantum dots in bulk media. However, we find a low contrast of 0.04 in the Ramsey interference. We identify as a main mechanism a significant spin-flip rate due to excess free charge carriers induced by the strong rotation laser. These effects could be mitigated on optimized noise-free devices or by coupling the rotation laser via the waveguide mode to a chirally-coupled quantum dot. The demonstration of coherent optical control unleashes the full potential of waveguide-based spin-photon interfaces for quantum information applications.

\begin{acknowledgements}
We gratefully acknowledge financial support from the Danish National Research Foundation (Center of Excellence `Hy-Q', grant number DNRF139), the European Research Council (ERC Advanced Grant `SCALE'), Innovation Fund Denmark (Quantum Innovation Center `Qubiz'), and the Danish Research Infrastructure Grant (`QUANTECH'). I.S., M.C.L. and R.J.W. acknowledge support from SNF (project 200020\_156637) and NCCR QSIT. R.S., A.L. and A.D.W. gratefully acknowledge support of BMBF (Q.Link.X 16KIS0867), the DFG (TRR 160), and DFG project 383065199. This project has received funding from the European Union's Horizon 2020 research and innovation programme under the Marie Sk\l{}odowska-Curie grant agreement no. 747866 (EPPIC) and no. 753067 (OPHOCS).
\end{acknowledgements}

%

\end{document}